\documentclass[11pt]{article}

\usepackage{amsfonts}
\usepackage{amsmath}
\usepackage{graphicx}% Include figure files
\newtheorem{assunzione}{Assumption}
\newtheorem{teorema}{Theorem}
\newcounter{Commento}
\newenvironment{commento}[1][]%
               { \addtocounter{Commento}{1}
                 \ifx\relax#1\relax
                 \blankm\noindent
                 \textbf{Remark~\arabic{Commento}.}\quad%
                 \else
                 \blankm\noindent
                 \textbf{Remark~\arabic{Commento}. {#1}}\quad%
                 \fi
               }%
               {\blank}%%
\renewcommand{\phi}{{\varphi}}

\newcommand{\vett}[1]{\mathbf{#1}} 
\newcommand{\blankm}{\vskip 0.5em}
\newcommand{\blank}{\vskip 1.0em}
\title{Spectra   as a classical phenomenon, 
and the Einstein classical  program 
}

\author{A.~Carati\thanks{Dip. di Matematica, Universit\`a degli studi
  di Milano, Via Saldini 50, I--20133 Milano, Italy, and Corso di
  Laurea in Fisica, same University} \and
  L.~Galgani\footnotemark[1] \and
  F.~Gangemi\thanks{Independent researcher.}}

\date{\today}

\begin{document}

\maketitle
\begin{abstract} 
  According to Born (\emph{Atomic Physics, page 103}), spectra are
  \emph{``quantum phenomena, which from a classical standpoint are
  perfectly unintelligible''}. However we illustrate results on
  classical calculations of infrared spectra of ionic crystals
  (actually LiF) which show that the situation is much more
  complex. Indeed it turns out that: 1) At room temperature and at
  higher ones (up to 1060 K) the
  classical computations reproduce the experimental data, even better
  than the \emph{presently available} quantum ones do; 2) At lower
  temperatures (even at 7.5 K), the classical
  computations reproduce pretty well the data, if one accepts the idea
  advanced in 1916 by Nernst (the inventor of the third principle)
  that zero-point energy has room in classical physics too.  It is
  eventually pointed out that the mentioned results might be regarded
  as a first step towards an implementation of the Einstein Classical
  Program, which aims at deducing quantum physics (admittedly the
  correct theory) from a realistic theory. In fact, we are considering
  the Einstein classical program in the extreme version in which the
  realistic theory is just (\emph{essentially, see below}) classical
  electrodynamics of matter in bulk, involving phase space orbits,
  solutions of Newton equations. An Appendix is devoted to
  illustrate the Nernst approach, which concerns also the relation between
  equipartition and Planck's law.
\end{abstract}  

\section{Introduction: The FPU (or FPUT) problem, and the Einstein Classical
  Program } 

The present paper is written for an issue of the Journal of
Statistical Physics devoted to the so-called FPU problem.  Now, about
70 years after such a problem was posed, one can quite naturally ask
which is, in the end, the significance of that problem.  Is one
dealing with a mathematical problem or with a physical one?  The key
contribution of this work consists of exhibiting that the FPU  problem
has a quite relevant physical significance for the general
\emph{question of the relationship between quantum and classical
physics}.

This is shown by just passing from studying the original Fermi
one-dimensional toy model \cite{fpu} (see also \cite{chaos}) to
studying a realistic classical model of a crystal of the type
currently used in solid state physics, and considering the possibility
of calculating spectra, which are classically perfectly unintelligible
according to Born (see \cite{born}, page 103). Indeed we
illustrate results obtained in recent years, which exhibit that, in a
classical framework, first of all the spectra of crystals can be
calculated (actually, in the case of the infrared spectra of the ionic
crystal LiF).  Concerning the comparison with experimental data, it
turns out that at room temperature and above it the agreement is
pretty good. For lower temperatures (even at 7.5 K) a classical
agreement is again obtained, but only if the assumption is made that a
zero-point energy has room in a classical framework too (see
later). Concerning quantum computations, they were performed
essentially within the Maradudin group. A complete calculation of the
infrared spectrum is apparently available only at room temperature,
and the agreement with experimental data is pretty worse with respect
to the classical case.  However, it will be seen that the difficulties
met in the quantum approach are apparently just of a technical type,
so that we are confident that the quantum calculations too will be
shown to be at the level of the classical ones.

Anyhow, the present situation is rather strange, since it seems that
(at room temperature) one has to establish whether the typical quantum
phenomenon of spectra, which is classically unintelligible according to
Born, is better reproduced by a quantum calculation or by a classical
one. Just in order to clarify our scientific position, we bet the
result will be fifty-fifty, two essentially equally good
reproductions.

And why? Because, we confess, our scientific position is that of
trying to vindicate the Einstein's Classical Program. As pointed out
by him in many occasions, and illustrated in a particularly vivid way
in the book \cite{schillp} \emph{Einstein Phylosopher Scientist} from the year
1949,\footnote{Edited in the occasion of his $75^{th}$ birthday, six
years before his passing away.}  (last chapter, \emph{Reply to
criticisms}), such a program consists of trying to prove that quantum
physics (admittedly the correct theory) can be deduced as a theorem
from a \emph{realistic theory}.  In fact, we assume such a program in
the extreme form in which the realistic theory is just classical
physics, which actually is the choice that Einstein himself did in his
paper with Stern from the year 1913.  More precisely, the extreme
version of the program we consider consists of assuming, as primitive
theory, classical electrodynamics of matter in bulk, extended as to
include the Dirac radiation reaction force from 1938 (see \cite{dirac38}), with
its fundamental qualitative complement, i.e., \emph{a global
nonrunaway asymptotic condition for $t\to+\infty$}.

However, in the specific case of spectra we are considering here, a
first step can be performed in the much more limited, purely
mechanical, framework, in which the radiation force is not taken into
account, and the electromagnetic field enters only in the non-retarded
approximation. Which is precisely the procedure followed within the
Maradudin school in a quantum framework. It is at such a purely
mechanical level that the comparison of the quantum and of the
classical results is performed in this paper. The very interesting
problem of implementing the Einstein classical program when
retardation is taken into account and a macroscopic field shows up in
the medium, is beyond the aims of the present paper, and only some
comments will be added later. 

Anyhow, even at the level of the present paper a relevant problem
remains open.  The point is that the classical computations of spectra
were performed in a state of equipartition, since a study of the ergodic
properties shows that, in the classical case, equipartition is
achieved in a short time. Conversely, the quantum computations à la
Maradudin were performed using the Planck distribution, of course. So,
how is it possible that there might be this compatibility between two
seemingly incompatible distributions, equipartition and Planck's law?
Now, this is precisely the problem studied from 1916 by Nernst (the
inventor of the third principle, and organizer of the first Solvay
conference), who showed that such a compatibility occurs if a suitable
coexistence of ordered and disordered motions is met. Such a solution
by Nernst, essentially unknown, is also illustrated in an Appendix.

In Section 2 existence and comparison of classical and quantum spectra
are exhibited, in Section 3 the quantum and the corresponding
classical model are described, Section 4 is devoted to the temperature
dependence of spectra and in Section 5 the ergodic properties of the
classical model are discussed. The conclusions follow. An Appendix
is devoted to the Nernst contribution.

\section{An apparent paradox: classical spectra exist}

It will be seen that, both in the quantum and in the classical
treatments discussed in this paper, spectra are calculated at a
purely mechanical level, without ever referring to the electromagnetic
field and to its propagation inside the crystal.  So, in the present
work devoted to spectra we will proceed at a purely mechanical level,
implementing Kubo's method, along the same lines followed in the
quantum framework within the Maradudin group.

Both in the classical and in the quantum cases the spectra were computed  in terms
of \emph{susceptibility} $\chi$. This is a thermodynamic
quantity\footnote{Indeed, the position of spectral lines in
gases depends, for example, on pressure. There is a vast literature on this phenomenon in
astrophysics, due to the importance that spectral lines have in the
study of stars. Regarding laboratory measurements,
we mention only the excellent doctoral thesis \cite{seldam} on the
variation of the polarizability of argon with pressure, and the
literature cited therein.}
which involves the time evolution of a macroscopic quantity,
\emph{polarization} $\vett P$, which in turn shows up as a mean of the
microscopic quantities $e_i \vett x_i$ involving positions and charges
of the single atoms. Thus no reference at all is made to microscopic
energy levels and possible jumps among them, which were the bases of
the old approach, that turned out to be ineffective. Indeed, at least
in the paradigmatic case of LiF, spectra are pure
thermodynamic quantities essentially unrelated to atomic properties,
other than positions and values of the charges.
Moreover, it is known that, in the infrared region, the
microscopic formula of susceptibility essentially depends only on
contributions from the motions of ions, since the contributions of the
electrons can be accounted for by an additive constant.

According to the Green-Kubo theory, as discussed for example in
\cite{cg}, the electric susceptibility can be derived microscopically
at a purely mechanical level in the following way. First one defines
the polarization per unit volume as
\begin{equation}\label{pol}
\vett P(t)= \frac 1V \sum_a e_a\vett x_{a}(t) \ .
\end{equation}
where ${\vett x}_a$ ($a=1, 2, \cdots $) are the positions of the
charges (actually ions) belonging to a ``small'' volume\footnote{The
volume $V$ must be small on a macroscopic scale, but large enough on a
microscopic scale. In our case $V$ is just the volume of the
simulation cell.}  $V$, while $e_a$ are the corresponding charges.
Then the susceptibility $\chi_{ij}(\omega)$, is given by the formula
\begin{equation}\label{eq:chi}
\chi_{ij}(\omega) = \frac V{K}\int_0^{+\infty} e^{-i\omega t} \langle
P_i(t) \dot P_j(0)\rangle d t \ ,
\end{equation}
where $K$ is the \emph{specific} kinetic energy of the ions, angle brackets
denote an average (a phase space average in the quantum case, and
essentially a time average - see below - in the classical case), and
$P_i(t)$ is the $i_{th}$ component of polarization at time
$t$.\footnote{Notice that for an isotropic crystal, as is the case of
LiF, the tensor $\chi_{ij}$ is isotropic and reduces to a scalar.}
\begin{figure}[th]
\includegraphics[width=0.5\textwidth]{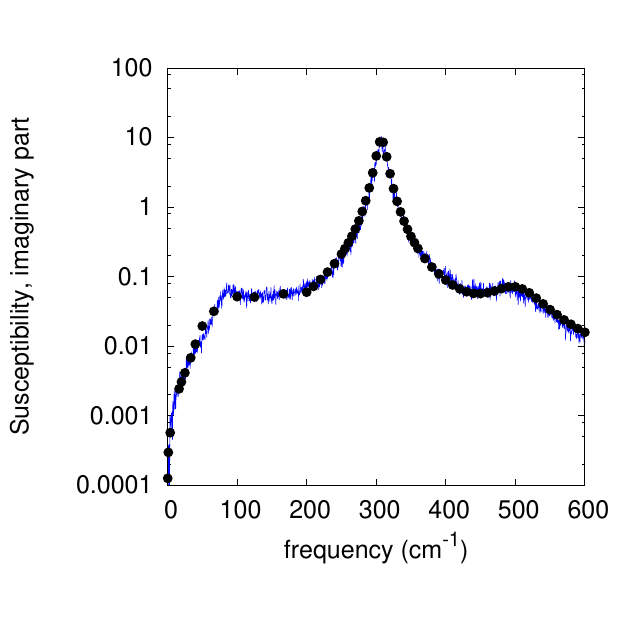}
\includegraphics[width=0.4\textwidth]{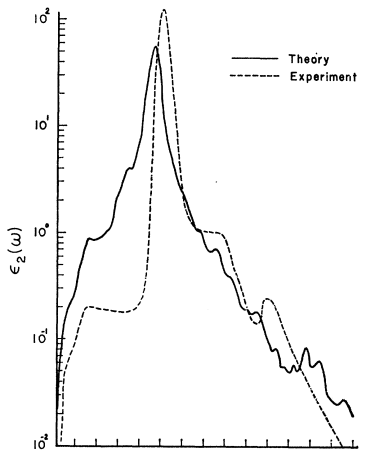}
\caption{Imaginary part of the electric susceptibility at room
temperature. Comparison of calculated and experimental values in a
  classical (left) and quantum (right) calculation.
\label{fig:1}}
\end{figure}

The best available classical and quantum results at room temperature
are reported in Fig.~\ref{fig:1} (left and right, respectively) where the
calculated imaginary part of the electric susceptibility is compared
with experimental data at room temperature.  In the classical case
(left panel, taken from Figure 1 of \cite{Li2}) an apparently very
good agreement is found on a wide range of infrared frequencies, and,
in addition to the main peak, two shoulders, due to anharmonic
contributions, are correctly reproduced. Furthermore, as shown in
Figure 2 of \cite{Li2} not reported here, the exponential decay at
high frequencies is also correctly reproduced.
Such decay simply follows from (\ref{eq:chi}) noticing
that classical trajectories are analytic as functions of time. More
difficult is to explain the exponential decay in quantum terms.
In any case the approximation of spectra, when performed
in terms of Lorentzian functions as made for example in paper
\cite{jasperse}, fail to capture this behaviour. 

The quantum computation (right panel) performed within the Maradudin
group, is taken from Figure~12 of paper ~\cite{marad} by Ipatova et
al.  It is likely that the poorer accuracy of the quantum computation
is due to the approximations that had to be performed, as explained in
more detail below.

\section{The model, and the calculation method}

In order to understand the differences between our calculations and
those of the Maradudin group some comments are needed about the
microscopic mechanical models adopted and the computational methods
employed.
  
First of all, the classical and the quantum models for the ions were essentially of
the same type, first proposed by Born. First of all, the ions were
dealt with as point particles, since the role of the electrons
(typically, a repulsive role at short distances), was taken into
account through a suitable short-range interionic potential.  This
however can be done at the cost of introducing a suitable effective
charge for the ions. In the quantum paper \cite{marad} the model of
Hardy~\cite{hardy,karohardy} was adopted, where, in addition to
electrostatic interactions, a short-range potential of the form
$a\exp{-(r/\rho)}$ was introduced, with parameters $a$ and $\rho$
derived from experimental data \emph{on nearest-neighbor distances and
compressibility}. Moreover, one more parameter was introduced in order
to take the role of polarizability into account. In our investigations
we use essentially the same model, but with a short-range potential of
the Buckingham type~\cite{buck} (often used in molecular dynamics
simulations of crystals). Such a potential has the form
$A\exp(-Br)-C/r^6$, where $r$ is the interionic distance and $A,B,C$
are suitable parameters determined by fit to experimental data: namely,
data on dispersion relations, and \emph{not} on spectra, were used for the
fit, where the free parameters were an effective charge of ions (the
same for all ions, except for the sign) and a different set of $A,B,C$
values for each one of the pair types Li--Li, Li--F, F--F.

Concerning the numerical method, in our works \cite{Li} and \cite{Li2}
the time-correlation function appearing in equation (\ref{eq:chi}) was
determined by means of molecular dynamics simulations of a system of
4096 ions contained in a simulation box, with periodic boundary
conditions, at given volume and energy. Instead of phase averages,
time averages were used, with additional averaging over multiple runs
starting from independent initial conditions, so that the aggregated
simulation time was of the order of nanoseconds. It is important to
specify how the initial conditions were chosen: in each run the
crystal was set in its equilibrium configuration and the atoms were
assigned random velocities, extracted from a Maxwellian distribution
at a given specific energy (different runs corresponding to different
values of these random velocities). The way in which temperature was
determined will be discussed in the next Section. With such choice,
\emph{the energies of the normal modes of the crystal are essentially
in equipartition from the beginning}. This fact might appear strange,
since we are going to obtain a typical quantum phenomenon (spectra) by
performing averages according to the most characteristic classical
measure. In any case, it turns out (see Section~\ref{subsec:equip}) out that in
our model too, as in the standard FPU one, equipartition is attained
even if one starts from an atypical situation with a few normal modes
initially excited.

On the other hand, the calculation method adopted in the quantum paper
\cite{marad} involves many approximations at various stages. The key
point is that in a quantum framework the calculation of the time
correlation of polarization which appears in formula (\ref{eq:chi})
(see \cite{mw1962}) was performed in terms of quantized normal mode
coordinates, and this entails that such energies are in a Planck
state. However the calculations requires a Taylor expansion of the Hamiltonian as
a function of normal-mode coordinates, extended to fourth order in the
most accurate calculations, and also the neglecting of the Coulomb
contribution to anharmonic terms, and several additional
simplifications as specified in \cite{marad}. We believe that these
approximations might explain the lower accuracy of the results
obtained in the quantum computation. Notwithstanding extensive
research, we couldn't find in the literature any paper giving results
more accurate than those by Maradudin's group, reported
above. Probably better results could be obtained at lower
temperatures, but no result seems to be available.
\begin{center}
\begin{figure}[t]
\includegraphics[width=\textwidth]{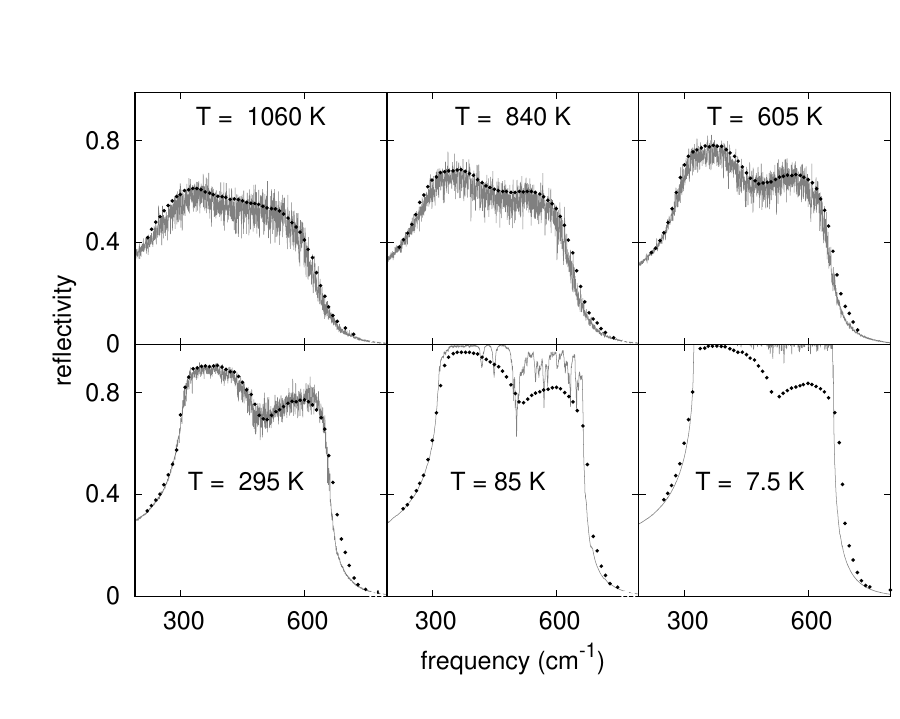}
\caption{Reflectivity
  %(a quantity related to susceptibility)
  computed classically for several temperatures (continuous lines), compared
  with experimental data (black dots). Note the discrepancy at the
  lower temperatures of 85 K and 7.5 K, but see
  figure~\ref{fig:4}, \label{fig:3}}
\end{figure}
\end{center}

\section {Dependence on temperature}
 
Experimental data on the infrared spectra of LiF are available in a
wide range of temperatures from a few kelvins up to almost the melting
point, as shown for example in \cite{jasperse}, where data on
reflectivity are reported for a range of temperatures from 7.5 K to
1060 K.

In Statistical Mechanics, temperature is usually introduced through
the Gibbs ensemble that is used in performing phase averages. In our
case, instead, averages were performed as time-averages, which is in
principle a better procedure.\footnote{See the discussion between
Einstein and Poincar\'e at the first Solvay conference, after the
relation given by Einstein.} In fact, in the case of crystals, the
system does not wander on the whole energy surface, but remains close
to the initial position.\footnote{For example, the ions do not
exchange their positions as would be allowed according to energy
conservation. In the words of Fermi, in a crystal particles are
distinguishable} So, it seems that the use of the Gibbs distribution
and of the corresponding identification of temperature might not be
justified, at variance with the case of gases.
\begin{center}
\begin{figure}[t]
\includegraphics[width=\textwidth]{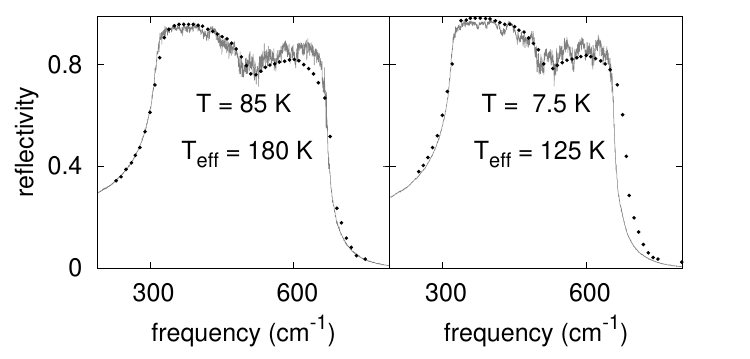}
\caption{Reflectivity computed at 180 K (left panel, continuous line)
  and at 125 K (right panel, continuous line) compared with
  experimental data at 85 K (black dots, left panel) and 7.5 K (black
  dots, right panel).
\label{fig:4}}
\end{figure}
\end{center}

So we had to investigate anew the relationship between temperature and
specific energy of the crystal (total energy divided by the number
of degrees of freedom).\footnote{In the following, the values of
specific energy will be given in kelvin $\mathrm{K}$ units, i.e.,
specific energy divided by the Boltzmann constant.}  The calculation
of spectra at different temperatures and their comparison with the
data available from reference \cite{jasperse}, was reported by our
group in \cite{Li2}, and is reproduced here in figures~\ref{fig:3} and
\ref{fig:4}.  At first, we took temperature as proportional to the
average kinetic energy $K$ of atoms, according to the usual relation
$K=3/2k_BT$, where $k_B$ is the Boltzmann constant. As one can see
from Figure~\ref{fig:3}, taken from \cite{Li2}, at high temperatures
the results are in good agreement with the data, whereas qualitatively
wrong results are obtained at low temperatures, especially at the
lowest one investigated, 7.5 K.

Note, however, that the Debye temperature for LiF is estimated to be
735 K, meaning that the crystal would be in the quantum regime even at
room temperature, where the classical computations are in agreement
with experimental data. Therefore, we tried to verify whether the
agreement could also be found at lower temperatures.

In fact, it was found that even low-temperature spectra can be
reproduced with good accuracy if simulations are run at suitable
higher energies: namely, in order to have good agreement with data at
85 K, simulations had to be run with a specific energy of 180 K, and
for data at 7.5 K a specific energy of 125 K was found to be
appropriate (see Figure~\ref{fig:4} also taken from \cite{Li2}).

So, it turns out that, for a crystal as LiF, the relationship between
specific energy and temperature is by far different from from the
direct proportionality usually assumed in molecular dynamics
simulations: indeed spectra can be reproduced classically even at
low temperatures if one assumes that kinetic energy doesn't vanish at
zero (absolute) temperature. In fact this was already interpreted in
the paper \cite{GGCG} as corresponding to the existence of a
zero--point energy\footnote{In German Nullpunktenergie.} in a
classical framework, along the lines proposed by Cercignani in the
paper \cite{cercgalscot}.

We could not yet determine analytically the relation between specific
energy and temperature, but it appears from the results on spectra
that it has a behaviour qualitatively similar to the quantum one, with
a zero-point energy that can be estimated around 120 K. We are not
aware of quantum computations of spectra for LiF at low temperatures.
%
% The FPU phenomenon
%
%
\begin{center}
\begin{figure}[t]
\includegraphics[width=0.8\textwidth]{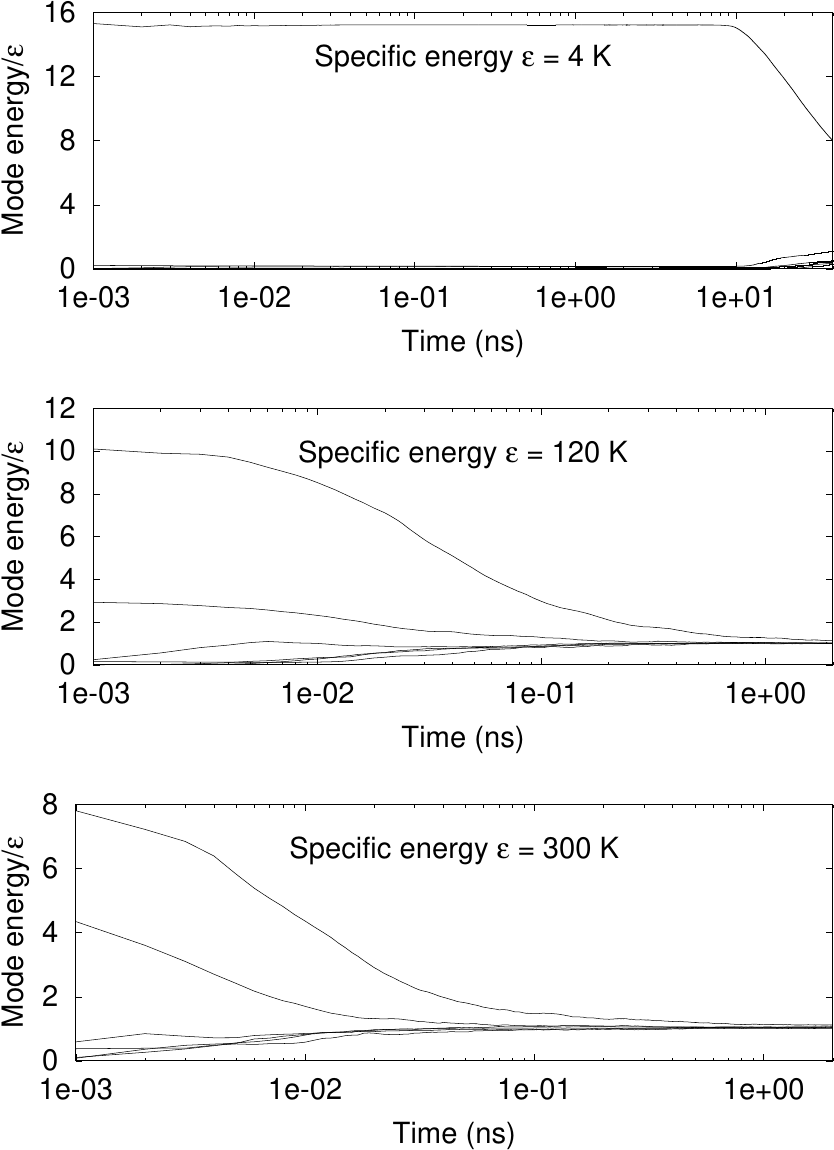}
\caption{Time needed to attain equipartition. Time averages of some
  mode packets energies (normalized by the specific energy
  $\varepsilon$) vs. time, for $\varepsilon=4$ K (upper panel),
  $\varepsilon=120$ K (central panel), and $\varepsilon=300$ K (lower
  panel).
\label{fig:5}}
\end{figure}
\end{center}

\section{Ergodic properties of the model}
\label{subsec:equip}

It was already mentioned that the time averages used to calculate
spectra in a classical framework were obtained by extracting the
initial data from a Maxwell-Boltzmann distribution (at a given
specific energy), that is, in a state of energy equipartition. In
fact, we were forced to choose an equipartition state, because after a
temperature-dependent short time (see below), the normal modes are
found to be in such state.

On the other hand, in the quantum case the results of the Maradudin
group were obtained by calculating the averages according to the
Planck distribution.  So, the strange fact seems to occur that energy
equipartition might be compatible with Planck's law. This is precisely
the problem discussed in 1916 by Nernst, the discoverer of the Third
Law and organizer of the first Solvay Conference. Such a work will be
illustrated in an Appendix, pointing out that compatibility is
achieved if one assumes that the system's energy can be thought of as
divided into an "ordered" part and a "disordered" one, and that the
"ordered" part plays the role of a sort of classical zero-point energy
reserve.

In this connection, we now illustrate how the spectra give a qualitatively
relevant information on the dynamics of our system, which could help
understanding the possible coexistence   of ordered and disordered motions.
In fact, as the polarization is given by a
linear combination of the two transversal optical normal modes of
vanishing wave vector, then the strong peak in the spectra exhibited in
Figure~\ref{fig:1}, means that the autocorrelations of such  two modes
keep oscillating for a long time without vanishing, implying that
their motion is partially ordered and partially chaotic. Fermi
himself was well aware that the ergodic properties of dynamics have a
significant influence on the physical properties of a system. Indeed,
he himself described the result of his work \cite{fpu}, in Ulam's
words, as ``\emph{a little discovery, in providing intimations that
the prevailing beliefs in the universality of ``mixing and
thermalization'' in nonlinear systems may not always be justified}''.

In fact, as is well known, in the FPU paper it was shown that,
starting from an atypical state, in which only a few normal modes are
initially excited, the system remained for long times in a state close
to the initial one instead of wandering on the entire energy surface.
On the other hand, the occurring of ordered motions happens also if
one starts from a typical condition.  For example, a study of an
FPU-type model with an initial equipartition state\footnote{More
precisely, within the microcanonical ensemble.} was performed in the
work \cite{cggp}. In that case, the autocorrelations of the
mode-energies were studied, and it was found that the high frequency
modes remain correlated, while the correlations of the low frequency
ones vanish. So one has strong indications that the motions remain
partially ordered also for generic initial data.  Notice that such a
type of computations are the counterpart of our numerical work on the
spectra.

It is then interesting to implement the original Fermi approach
recalled above, also in the case of Lithium Fluoride (for a different
way to explore the ergodic properties of the LiF see
paper~\cite{physA2019}). To this end, the normal modes of a cell with
64 atoms were first determined and grouped together according to
their degeneracy. Then numerical simulations were run with all the
energy initially given to the lowest-frequency group, for a given
specific energy $\varepsilon$. In particular, three cases were
considered: $\varepsilon =120$ K, corresponding to our estimate of the
zero-point energy, and two more values, below and above it
respectively, actually 4 K and 300 K. For each group $j$, the
instantaneous mean energy, $E_j(t)$, and the time-averaged energy,
$\bar E_j(t)=1/t\int_0^tE_j(t')dt'$, were calculated during numerical
simulations. The results are shown in Figure \ref{fig:5}, where the
ratio $\bar E_j(t)/\varepsilon$ is reported for some selected groups
of modes: equipartition is attained when $\bar E_j(t)/\varepsilon=1$
for each group of modes. At a specific energy of 4 K (well below the
zero-point energy, estimated of about 120 K in our units) the
formation of a metastable state in which the packet energies are
approximately constant can be observed for a time interval of the
order of 1 nanosecond, after which there is a gradual transition to
equipartition (consider the logarithmic scale on the time
axis). Instead, at a specific energy of 120 K the system immediately
starts evolving towards equipartition, and no metastable state seems
to occur. An even faster approach to equipartition is observed at 300
K. Thus one may conclude that for any physically significant state of
the crystal (a state with specific energy above zero-point energy)
equipartition occurs in a very short time.

By such a result, one could then conjecture that the system behaves in
an ergodic way (on a short time--scale). On the other hand, the fact
that the spectrum has a strong peak about a certain frequency exhibits
that the motion of the considered system is not completely chaotic,
and presents a certain degree of order.

A way to reconcile such an apparent discrepancy,
might be the one suggested by Benettin, Ponno and Christodoulidi in the works
\cite{benponno1,benponno2}, that were performed in order to explain
the showing up of a metastable state in the FPU--$\alpha$ model. The
idea is that a better description of the FPU-like dynamics should be
obtained thinking of it not in terms of normal modes, but as a
perturbation of an integrable Toda system. The quantities that remain
constant for very large times should be the actions of the Toda
system, and not the normal modes. For small specific energies, the
normal modes are a small deformation of the Toda actions so that they
too turn out to be almost constant, whereas at higher energies the
deformation increases and so a normal-modes packet is formed that
presents a local equipartition, and such a packet may even lead to an
apparent equipartition for all modes.

An analogous phenomenon might occur also in the case of Lithium
Floride, namely there could be an integrable structure, in terms of
which the system is better described, with respect to the description
in terms of normal modes. In the Appendix illustrating the Nernst work
from 1916, we will show how the presence of a certain degree of
``ordered motion'' might affect the value of relevant thermodynamic
quantities. In particular we will show how Nernst conceived of a
compatibility between equipartition of the mechanical mode energies
and Planck's law for the thermodynamic energy.
 
\section{Conclusions}
\label{sec:concl}

So we exhibited that, against the sharp opinion of Born, spectra can be
computed classically with a mechanical model, in the approximation in
which the role of the electromagnetic field is completely neglected,
as also was done in all the available quantum computations. In any
case, it is a fact that, presently, the classical results turn out to
be even better than the quantum ones. Now, it is true that the
difficulties with the quantum computations seem to be essentially of a
technical type, so that we are confident that the quantum computations
at room temperature may be found to agree with experimental data as
well as the classical ones do. However, it is also a fact that the
present situation somehow seems to constitute a challenge for quantum
mechanics.

Is it possible that all this happened by chance? We are convinced that
this is not the case, and that we are rather meeting here with a
concrete first step of an implementation of \emph{the Einstein
classical program}, which consists of trying to deduce quantum
mechanics (admittedly the correct theory) from a ``realistic theory'',
possibly just from classical physics. Now, the results illustrated
here in connection with spectra, were obtained in the mechanical
approximation in which the role of the electromagnetic field is
neglected.  However, a complete implementation of Einstein's program
in the extreme form which includes the electromagnetic field, was
actually performed in the work \cite{alessio}, albeit only for a
\emph{linearized} model of a classical crystal. Indeed it was there
shown that in the crystal the \emph{macroscopic} electromagnetic field
propagates with a velocity different from $c$, and moreover that
polaritonic branches appear in the dispersion relations. To our
knowledge, such a result was not yet achieved in quantum
electrodynamics.
Some other results towards an implementation of the Einstein classical
program were already obtained.  See for example the
works~\cite{bell,coppie,ioneH+} and the book~\cite{librocg} in
preparation.

Concerning the problem of recovering quantum mechanics in a classical
framework, the Nernst attitude was obviously the same of Einstein.
And apparently this also occurs with Dirac: see for example the
chapter devoted to Dirac in the book \cite{bokulich}.  But even
more. We are convinced that this was the attitude of Fermi too. Indeed
this makes understandable why Fermi considered a ``little discovery''
the first results obtained in his toy model, as indicating that
systems of physical interest may not be ergodic. In any case, this was
confirmed to the senior author during a long conversation he had in
Berkeley with E.~Segr\`e, the Fermi's pupil, also the editor of
Fermi's ``Collected Papers''. In such a conversation, Segr\`e recalled
how Fermi, due to his shy character, disliked publicly discussing
problems on the foundations of quantum mechanics, which could lead to
pseudo--philosophical discussions. However, Segr\`e vividly recalled
how, when discussing with his pupils, Fermi explicitly stated that he
had, regarding the relationship between classical and quantum
mechanics, a position similar to that of Einstein. Among Fermi's
pupils, albeit at a different level, was also Piero Caldirola, the
leader of the theoretical physicists group at the Milan University in
the years 1960's and 1970's, he too had the same recollection of
Segr\`e. And this explains why the first works on the FPU problem
\emph{oriented in the spirit of the Einstein classical program}, were
performed at Milan: first of all the Bocchieri, Scotti\footnote{The
master of the senior author.}, Loinger paper from 1970 (see
\cite{BSL}), soon followed by the Galgani, Scotti work \cite{GS} from
1971, in which it was suggested that even Planck's law might have room
in the FPU problem, and shortly later followed by the paper \cite{CGS}
from 1972 in which Cercignani (with Galgani and Scotti) proposed that
zero-point energy too might have room in the FPU problem. This in fact
promoted a kind or rebirth of the FPU problem.

\vskip.5 truecm \textbf{Dedication}. This paper is dedicated to the
memory of Giuseppe Pastori Parravicini, an italian authority in solid
state physics, schoolmate of the senior author at the Milan
University. It occurred to him to come to know of a result of the first
two authors obtained in 2003 \cite{CG2003}, in which a feature of
electrodynamics of matter in bulk was pointed out, that had escaped
Born and Huang \cite{bornhuang} and possibly the whole community of
solid state physicists. Then he pointed out that through such a result
it had become possible to deal within \emph{purely microscopic} models
of matter in bulk involving retardation, proceeding as if quantum
physics did not exist. So, phenomena such as the existence of
polaritons and propagation of the macroscopic electromagnetic field in
matter with a velocity different from $c$, were proven in a purely
classical framework. This opened within our group the road to the
results illustrated here, and to several others.

\appendix
\section*{Appendix: The Nernst theory from 1916}
\label{sec:comp}

\subsection*{The Nernst idea of a stochasticity threshold, and
  its implementation }
\label{nernst}

So we have seen that spectra, one of the most typical quantum phenomena,
were calculated classically in a state of energy equipartition. But
this seems paradoxical, because quantum mechanics itself had origin
through Planck's law, as opposed to equipartition. So it seems that
equipartition should somehow be compatible with Planck's law. How can
be this possible, and how can a statistical mechanics be formulated in
such a situation? It was mentioned in the Introduction that already in
the year 1916, stimulated by the works of Poincar\'e (1912) and of
Einstein-Stern (1913) on the relationship between Planck's law and
equipartition, Nernst (the inventor of the third principle, and
organizer of the Solvay conference of 1911), in the paper~\cite{nernst}
advanced a very peculiar theory in which Planck's law is deduced in a
completely classical framework, in an equipartition state, under a
suitable dynamical conjecture. We illustrate here the way in which,
fulfilling a reinterpretation of the Nernst ideas initiated almost 50
years ago (see papers~\cite{luiginernst1,luiginernst2,luiginernst3}),
we understand the Nernst theory.

Such a theory is based on the two  assumptions given below.
\begin{assunzione}
The oscillators of any given frequency $\nu$ are thought of as
subdivided into two groups according to their instantaneous energies
$E$: the ``ordered'' ones having energy $E<h\nu$, and the
``disordered'' or chaotic ones having energy $E>h\nu$. Furthermore,
their energies are assumed to be Maxwell--Boltzmann distributed.
\end{assunzione}  
In such a way there remains defined \emph{the conditional mean
specific energies $U_1$ and $U_0$ of the ``disordered'' (above the threshold)
and of the ``ordered''} ones respectively, and one finds (see the
proof in the next subsection)
\begin{equation}\label{eq:3}
  \begin{split}
    U_1&=h\nu + k_B T \ \\ 
    U_0&=k_BT- U_{pl}  \qquad   \mbox{i.e.,\quad  $k_BT= U_0+U_{pl}$} \ ,
  \end{split}
\end{equation}
where $ U_{pl}$ is the Planck energy
\begin{equation*}
U_{pl}=\frac  {h\nu} {\exp (h\nu/k_BT)-1} \ ,
\end{equation*}
which unexpectedly shows up in a completely classical framework.

In particular the second of (\ref{eq:3}) in the form $k_B T=U_0+U_{pl}$
exhibits that the complement of Planck's energy to $k_BT$ is of
ordered type (in fact, just the mean value of the ordered energy). 

Such formulas are those needed for our reinterpretation of what Nernst
calls his \emph{Erster Weg} (first way) method. In fact, with respect to
Nernst we also added here the formula of $U_1$ that he does not
mention, but plays for us a relevant role, since it allows one to
avoid his whole \emph{Zweiter Weg} (second way), which is very
interesting, but not at all easy. The key role of $U_1$ is that it
allows one to introduce the following
\begin{assunzione} The \emph{thermodynamic energy} $U$ of the system
is given by what is sometimes called the \emph{disposable energy},
namely,
\begin{equation*}
U=U_1-U_0\ .
\end{equation*}
\end{assunzione}
This is indeed analogous to what is done in the liquid-gas
case.\footnote{We thank Moreno Meneghetti and Francesco Ancilotto of
the Padua University for useful discussions in this connection.} This
in fact is also the choice repeatedly proposed by Boltzmann, whose
preferred example was that of a perfectly smooth sphere. Its total
mechanical energy is the sum of the center-of-mass energy and of the
rotational one; but the latter is a constant of motion, so that the
thermodynamic or disposable energy is obtained by subtracting the
rotational energy from the total one. Analogous arguments were given
in a celebrated letter of Boltzmann \cite{boltz} to the journal Nature
in the year 1895. See also \cite{nature-bgg}. For a review about
Boltzmann in this connection see \cite{barbara}.

Thus the formulas (\ref{eq:3}) and assumption~2 allows one to
obtain
\begin{teorema}[Nernst classical deduction of Planck's law]
For Maxwell--Boltzmann distributed oscillators of frequency $\nu$, in
the presence of a stochasticity threshold $h\nu$, the thermodynamic
energy $U$ is given by
\begin{equation}
  U= h\nu + U_{pl} \ ,
  \end{equation}
i.e., Planck's law, endowed with a zero--point energy
$h\nu$.\footnote{The
Nernst value $h\nu$ is twice the familiar value, but actually
coincides with the value found in the year 1913 in the celebrated
Einstein-Stern paper in which they gave a classical deduction of Planck's law.}
\end{teorema}
\begin{commento}[Planck's formula without zero-point energy.]
  One can introduce the fraction $n_1$ of the oscillators above the
  threshold and the fraction $n_0=1-n_1$ of the oscillators below the
  threshold. Then one finds \cite{luiginernst3}
  \begin{equation}\label{eq:4}
    U_{pl}= n_1(U_1-U_0) \ .
  \end{equation}
  This immediately follows from $U_{pl}=k_BT-U_0$ (formula
  (\ref{eq:3})) by using the obvious formula
  \begin{equation*}
    n_0 U_0 + n_1 U_1 = k_B T \ .
  \end{equation*}
\end{commento}
\begin{commento}[Two zero--point energies?] The above zero--point
  energy $h\nu$ shows up in the formula for the thermodynamic energy
  $U$, as a constant additive term which does not influence for
  example the specific heat. But which relation does $h\nu$ have with
  the temperature dependent mean ordered energy $U_0(T)$ determined above,
  which is the complement of Planck's energy with respect to $k_BT$,
  and reduces to $h\nu/2$ for high temperatures? In fact such two
  energies are so much related that sometimes Nernst even calls them
  by the same name, zero--point energy. This was a difficult point for
  us. However, by a careful reading of his \emph{Erster Weg} and of a
  paper of Bennewitz and Simon \cite{bennewitz}, two pupils of Nernst, 
  one has the impression that, according to him, $h\nu$ acts as an
  ordered--energy \emph{reserve} parallel to the heat reserve of a
  calorimeter. So the system has two energy reserves: the heat of a
  calorimeter, and the ordered energy of the zero point reserve
  $h\nu$.
\end{commento}
\begin{commento}[Role of the Zweiter Weg.]
  We already mentioned that the Nernst proof is composed of two parts
  called \emph{Erster Weg} and \emph{Zweiter Weg} (First way and
  Second way). Above we reinterpreted the first way making it become
  very short, by a simple calculation of $U_1$, and by exploiting the
  Boltzmann--type prescription $U=U_1-U_0$. Thus there is no more need
  of the rather difficult \emph{Zweiter Weg} of Nernst. The latter
  however is very interesting, since it highlights that one is dealing
  here with a kind of chemical reaction in which a part of the
  zero-point energy reserve $h\nu$ goes to equilibrium with Planck's
  energy, becoming the ordered thermodynamic energy $U_0$.
\end{commento}
\begin{commento}[Mentions of the Nernst theory.]
 The Nernst theory is mentioned by Kragh (see for example
 \cite{kragh}), who however discusses the possible role of Nernst
 theory for cosmology, without discussing the theory. Furthermore,
 there are several papers by Boyer (see for example \cite{boyer,
   boyer2, boyer3}) concerned with a classical deduction of Planck's
 law that were very useful for us, since it is just through them that
 in Milan the paper of Nernst happened to be known in the years
 1970's. But the relation of such papers with the Nernst work is not
 clear to us. We hope to have the possibility to come back to this
 problem in the future.
\end{commento}

\subsection*{A compact proof of the Nernst  formulas }
\label{subsec:proof}

We come now to a proof of the two formulas (\ref{eq:3}) given above
for the conditional energies $U_1$, $U_0$. To this end we recall that
according to the MB distribution, in the usual unconditional case the
mean energy is given by
\begin{equation*}
  <E>_{MB}= -\partial_\beta \log Z_{cl}(\beta)
\end{equation*}
where the ``partition function'' is the familiar one,
\begin{equation*}
  Z_{cl}(\beta)= \int_0^{\infty} e^{-\beta E} dE \ .
\end{equation*}
So, for the ``ordered'' and the ``disordered'' cases ($E<h\nu$ and
$E>h\nu$ respectively), one evidently has the analogous formulas with
the corresponding integration intervals, i.e., the ``partial''
partition functions
\begin{equation*}
  Z_0=\int_0^{h\nu} e^{-\beta E} dE \ , \quad Z_{1}(\beta)=
  \int_{h\nu}^{\infty} e^{-\beta E} dE \ .
\end{equation*}
In the case of $Z_1$ the formula for $U_1$ is obtained immediately
through the change of variable $E=E'+h\nu$, because $e^{-\beta h\nu}$ factors
out, and through the logarithm the product is transformed into a sum.
In the ordered case, one has to look for a suitable factorization of
$Z_0$, and one more step is required. Preliminarily one should recall
that, as pointed out by Einstein in the first pages of his paper from
1907 on specific heats, Planck's formula is obtained through the
quantum partition function defined by
\begin{equation*}
  Z_{ein} = \sum_{n=0}^{\infty} e^{-\beta n h\nu} \ .
\end{equation*}

But it turns out that $Z_{ein}$ has a classical meaning too. Indeed,
this is seen by considering the classical partition function $Z_{cl}$,
and splitting the domain of integration into intervals of length
$h\nu$. Thus, for any $n>0$, one performs an analogous change of
variable $E=E'+nh\nu$ and one finds
\begin{equation*}
  Z_{cl}= Z_0 \cdot Z_{ein} \ ,
\end{equation*}
that immediately leads to the key formula $k_BT=U_0+U_{pl}$.

%\appendix{Einstein, Fermi and Nernst among the malcontents
%of quantum mechanics}
%%%

 \vskip 2 truecm
%
% The FPU phenomenon
%
%

\end{document}